\documentstyle[twocolumn,prl,aps]{revtex}
\large
\begin{document}
\draft
\twocolumn[
\hsize\textwidth\columnwidth\hsize\csname @twocolumnfalse\endcsname

\draft
\title{
       High frequency sound waves in vitreous silica
      }
\author{
        R.~Dell'Anna$^{1}$, 
        G.~Ruocco$^{1}$
        M.~Sampoli$^{2}$, and 
        G.~Viliani$^{3}$
        }
\address{
         $^1$
         Universit\`a di L'Aquila 
         and Istituto Nazionale di Fisica della Materia
         I-67100, L'Aquila, Italy. \\
         $^2$
         Universit\'a di Firenze
         and Istituto Nazionale di Fisica della Materia
         I-50139, Firenze, Italy.\\
         $^3$
         Universit\`a di Trento 
         and Istituto Nazionale di Fisica della Materia
         I-38050, Povo, Trento, Italy.
        }
\date{\today}
\maketitle
\begin{abstract}
We report a molecular dynamics simulation study of the sound waves in 
vitreous silica in the mesoscopic exchanged momentum range.
The calculated dynamical structure factors are in quantitative 
agreement with recent experimental inelastic neutron 
and x-ray scattering data.
The analysis of the longitudinal and transverse 
current spectra allows to discriminate 
between opposite interpretations of the existing experimental data in favour
of the propagating nature of the high frequency sound waves.
\end{abstract}
]
In the last $20$ years great effort has been devoted to understanding
the microscopic dynamics in topologically disordered systems.
The need originates from the experimental observation that a large
variety of glasses exhibits common dynamical and thermodynamical
properties, which are anomalous with respect to the corresponding crystals
\cite{Elliotlibro}. In particular, 
around $5$ K the thermal conductivity has a plateau and 
in the $10\div 30$ K temperature 
range the reduced specific heat $C_P(T)/T^3$ goes through a maximum, indicating
the existence of excess vibrations in the low-frequency 
range with respect to the Debye model.
In a more extended range, below $\approx 10$ meV, inelastic incoherent
neutron \cite{BZNGP88} and Raman \cite{WINT} 
scattering reveal a strong
broad band, usually called {\it boson peak}, 
which can again be related to an 
excess of the vibrational density of states. 
At the present, 
a unique microscopic description of 
these excess modes does not exist and their theoretical 
interpretations include localized vibrational states \cite{ORB92}, 
soft anharmonic vibrations \cite{BUCH2}, relaxational motions 
\cite{BZNGP88}, 
intrinsically diffusive \cite{AF93} and propagating 
\cite{grenoble} modes.
   
Among disordered systems,  vitreous silica is a 
typical strong glass former and has widely been 
studied, both experimentally \cite{BZNGP88,WINT,grenoble,BUCH3} and 
numerically \cite{Elliot2a}. The temperature analysis of the inelastic neutron-scattering
intensity \cite{BZNGP88} 
suggested the coexistence of sound waves, below $\approx 4$ meV, with a second class of local harmonic excitations, which become 
increasingly anharmonic towards the lowest frequencies. 
However, kinematic reasons did not (and do not yet) allow the use of 
the neutron scattering technique to directly study the acoustic
excitations in the boson peak energy region.
Only recently it became possible to utilize the 
inelastic x-ray scattering (IXS) with meV energy resolution \cite{ESRF}
to measure the dynamic structure factor, $S(Q,E)$, in the first 
pseudo-Brillouin zone. Using this technique, the high frequency dynamics of
v-SiO$_2$ at $T=1050$ K has been studied in the $1 \div 6$ nm$^{-1}$ 
momentum transfer range. The measured $S(Q,E)$ clearly indicate the existence of 
collective propagating excitations up to $Q=4$ nm$^{-1}$, i.e. up to
an energy of $\approx 13$ meV. The energy of these excitations extends both below and above
the boson peak region, located at $E\approx 4$ meV \cite{grenoble}.
Recently this  result has been confirmed in the whole 
$1800 \div 300$ K temperatures range
\cite{sil3}. At variance with this interpretation,
analogous room temperature IXS data combined with non-kinematic
inelastic neutron (INS) spectra \cite{courtens}
were interpreted as an indication of the existence  of a cross-over 
from sound waves to localized acoustic modes, which should be responsible 
for the boson peak.  
 
In order to clarify the propagating nature of low frequency modes around the
boson peak region, this letter presents a molecular dynamics (MD)
and normal mode analysis (NMA) study of the high frequency dynamics 
in vitreous SiO$_2$ in the $Q=1.4\div 5.1$ nm$^{-1}$ range. The reliability of our 
investigation is validated by the good agreement of the calculated $S(Q,E)$ 
with the available INS \cite{BUCH3} and IXS \cite{grenoble} data. 
The detailed knowledge of the spectral shape allows us to discriminate 
between the two proposed interpretations \cite{grenoble,courtens} and 
to establish the propagating nature of both 
longitudinal and transverse modes giving rise to the boson peak.
Moreover,
the widths of the excitations calculated in the harmonic approximation
(i.e. via NMA) are in quantitative agreement with both MD and experimental data,
indicating that in the examined $Q$ range their origin is 
structural rather than dynamical.
 
Standard microcanonical MD simulations with periodic boundary conditions
were carried out for systems 
consisting of $648$ and $5184$ particles enclosed 
in cubic boxes of length $L=2.139$ nm and 
$L= 4.278$ nm respectively (mass density $\rho =2.20 $ g/cm$^3$).
We used 
the two- and three-body interaction 
potential proposed by Vashista {\it et al.} 
\cite{VASH} which, compared with the simpler- 
to-deal two-body potentials, gives a better 
description of the structural properties 
of SiO$_2$. Details of this study will be discussed in 
a subsequent paper. The electrostatic long range interactions were taken 
into account by the tapered reaction field method and 
the equations of motion were integrated with leap-frog algorithm 
using a time step of $\Delta t = 0.5$ fs.
A $\beta$-cristobalite crystal was 
melted at $15,000$ K and equilibrated for a long time so that the initial
state has no effect on the final configuration. The system was
then cooled and thermalized through different temperatures ranging
from $5000$ K down to $300$ K. The temperature dependence 
of the total internal energy
and of the self-diffusion constants $D_{Si}$, $D_O$ are studied 
to recognise the structural arrest, which is found at  about $2700$ K, 
a value  definitely higher than the 
experimental glass-transition temperature 
\cite{ANG} ($T_g \approx 1450$ K).
This can be an effect, at least partially, of the quench rate \cite{VOLL}, which in our MD simulation is 
much higher
($\approx 10^{13}$ K/s) than the experimental value ($\approx 10^2$ K/s). 
Two different techniques were adopted to study the dynamic of the system in the glassy state:
{\it i)} (MD) the finite temperature ($T=1500$, $1000$, $600$, and $300$ K)
atomic dynamics has been followed for $60$ ps by 
recording the instantaneous configurations every $0.01$ ps;  
{\it ii)} (NMA) the dynamical matrix has been calculated in a 
"glassy" (metastable) energy minimum, reached by the steepest-descent 
method \cite{SW83}, and then 
either diagonalized to obtain a complete set of 
eigenvalues and eigenvectors or used in
the application of the spectral moments method \cite{BENOIT1,MOMENTITN}.

We investigated the dynamical correlation
in v-SiO$_2$ in terms of neutron- ($S_N({\bf Q}, E)$) and
x-ray- ($S_X({\bf Q}, E)$) weighted  dynamic structure factors. 
The presence of two distinct atomic species requires the calculation of the
partial dynamic structure factors
$$
S_{\alpha\beta}({\bf Q}, E) =
\int_{-\infty}^\infty
\frac{dt}{2\pi} e^{-i\omega t } 
\sum_{i \in \alpha}^{N_\alpha}
\sum_{j \in \beta}^{N_\beta}
\langle \frac{e^{i {\bf Q} \cdot {\bf r}_j(t)}}{\sqrt{N_\alpha}}
\frac{e^{-i {\bf Q} \cdot {\bf r}_i(0)}}{\sqrt{N_\beta}}
\rangle 
$$
where summation over $i$ and $j$ run over all atoms of type $\alpha$ and 
$\beta$ respectively 
and ${\bf r}_j(t)$ is the position vector of
atom $j$ at time $t$. $S_{\alpha\beta}({\bf Q}, E)$ is easily
evaluated from the MD trajectories. In the NMA one 
introduces the displacements $\{ {\bf u}_i\}$ from the equilibrium positions 
$\{ {\bf X}_i\}$ and expands these in terms of the normal modes; 
in the ${\bf Q}\cdot {\bf u}_i\rightarrow 0$ limit the {\it one excitation} 
contribution can be calculated as
\begin{eqnarray}
\label{eq1}
& &S_{\alpha\beta}({\bf Q}, E) = k_BT \sum_\lambda 
\frac{\delta(\omega - \omega_\lambda)}{\omega^2} \times \\
\nonumber
& &\sum_{i \in \alpha}^{N_\alpha} \sum_{j \in \beta}^{N_\beta}
\frac{e^{-W_{ij}}} {\sqrt{N_\alpha N_\beta m_\alpha m_\beta}}
{\bf Q}\cdot {\bf e}_i(\lambda)
{\bf Q}\cdot {\bf e}_j(\lambda)
e^{i {\bf Q} \cdot ({\bf X}_i - {\bf X}_j)}
\end{eqnarray}
where the Bose-Einstein factor has been approximated by its classical limit,
$W_{ij}$ is the Debye-Waller factor and 
$\{ \omega_\lambda \}$ and $\{ {\bf e}_i(\lambda) \}$  are the 
eigenvalues and eigenvectors of the dynamical matrix of the system.
$S_N(Q,E)$ and $S_X(Q,E)$ are then obtained by
adding the partial dynamic structure factors weighted by the atomic species 
concentrations and by the corresponding neutron or x-ray scattering lengths. 
An average over the independent directions of ${\bf Q}$, i.e.  
${\bf Q}=(\pm l,\pm m,\pm n)2\pi / L $, with $l, m, n$ 
integer numbers, is also performed.  
The reliability of the harmonic approximation is probed by the
good agreement existing between the $S({\bf Q}, E)$ calculated via NMA
and via MD at finite temperature. 
Moreover, the systems with $648$ and $5184$ atomes give identical
$S({\bf Q},E)$ in their common $Q$ range, indicating negligible size
effects.

Fig. $1$ compares in a very satisfactory way INS data (squares)
of ref.~\cite{BUCH3} measured at $5.4$ meV 
(corresponding to the position of the boson peak at $T \approx 1100$ K) 
with the neutron-weighted dynamic structure factor
$S_N(Q,E)$ calculated by our MD approach at the same temperature. 

We have also calculated the longitudinal current spectra $C_L(Q,E)$, 
defined as $C_L(Q,E)= E^2 S(Q,E)/Q^2$, at selected small $Q$ values 
($Q_{MIN} = 1.47 $ nm$^{-1}$ for the system with $L=42.78$ A). 
These spectra, reported in Fig. $2$ for some low-$Q$ values, 
have been fitted by the same model function (Damped Harmonic Oscillator, DHO) 
already used \cite{grenoble} to analyse the experimental spectral shape. 
In Fig. $3$ we report as full circles the best fit values of
the excitation energies $\Omega(Q)$, corresponding to the
positions of the current spectra maxima, and of the excitation widths 
$\Gamma(Q)$. 
The form of the current spectra
and the positions of the current spectra maxima are in qualitative agreement 
with those of \cite{kobvigo}.
The linear dispersion observed for $\Omega(Q)$ gives a
longitudinal sound velocity which is slightly higher than that
compatible with IXS data \cite{grenoble}; this could be due  to the thermal
history and quench rate of the numerical model, which can produce a tempering 
of the vitreous structure. The evident linear 
$Q$ dependence of the current spectra 
maxima positions, as already observable from the raw data of Fig.~2, indicates 
the propagating nature of the excitations up to $\approx 20$ meV, a  
range which includes the boson peak. 

This finding is not in contrast with earlier INS results by Buchenau et al. 
\cite{Buch4}, which demonstrate that pure plane-wave eigenvectors cannot explain 
the $Q$-dependence of the $S(Q,E)$ at $E\approx 4$ meV. Indeed, as it was shown
also in other simulated glasses \cite{mazza}, the eigenvectors of propagating 
modes present a random uncorrelated component superimposed to the plane-wave.

There is quantitative agreement between the $\Gamma(Q)$ 
values obtained from our simulation and that measured by IXS for
temperature ranging from $300$ to $1800$ K \cite{sil3}, both having 
a  $\Gamma(Q) \propto Q^2$ behavior.
The agreement of the $\Gamma(Q)$ values found in our {\it harmonic 
approximation} with those measured in the {\it real} and simulated system, 
strongly indicates
that the excitations linewidth, in the investigated $Q$ range, 
is likely to be related  to the structural disorder 
rather than to anharmonicity or other dynamical effects.
As suggested in ref.~\cite{grenoble}, the structural origin of the high 
frequency excitations widths is also supported by the temperature 
independence of $\Gamma(Q)$ derived from IXS data.
The picture becomes more complicated at small $Q$. 
Although the $\Gamma(Q) \sim Q^2$
law found at temperature above $\approx$ 300 K
by IXS and MD/NMA in the nm$^{-1}$ range,  quantitatively
extrapolates over two decades in exchanged momentum
down to the Brillouin light scattering (BLS)range ($Q \approx 0.02$ nm$^{-1}$),
suggesting a common structural origin of the $S(Q,\omega)$ linewidth,
at $T <$ 100 K the $\Gamma(Q)$ measured by BLS shows a strong
temperature dependence \cite{PV76}. 
Consequently, a structural origin of
$\Gamma(Q)$ at high $Q$ values is consistent with available data, 
while at small $Q$ the dynamical effects could be predominant. 
At present we have no explanation neither for such
different behaviour in different $Q$ ranges nor for the $\Gamma(Q) \sim Q^2$
law observed at $T >$300 K in the $Q=0.02 \div 5$ nm$^{-1}$ range.

In ref.~\cite{courtens}, the localized nature of the high frequency 
excitations was derived from the lineshape analysis of the IXS data.
However, the presence of a strong elastic contribution
and the poor statistics, 
did not allow to discriminate between
different lineshape models. 
A detailed lineshape analysis can be performed on
the simulated $C_L(Q,E)$ 
spectra. In the inset of Fig.~$4$ we report an example of 
fits of our calculated longitudinal current spectra with the expression 
proposed in ref.~\cite{courtens}.
Reliable fits can be obtained only leaving the important $\omega_{IR}$
parameter free. As reported in Fig.~4, the crossover frequency
$\omega_{IR}$, that
according to the localization model should indicate the localization edge 
and therefore the boson peak energy, is strongly $Q$-dependent 
and assumes values definitely higher than that of the boson peak. 
If we impose the same value of the $\omega_{IR}$ parameter for the different
$Q$ spectra, the fits result unacceptably bad.
 
To investigate the nature of the propagating excitations giving rise to 
the boson peak, we have also calculated the transverse current spectra $C_T(Q,E)$ 
at selected $Q$ values. The transverse spectra are not experimentally
accessible and are obtained from the MD via the correlation function of the
transverse current or from NMA substituting (${\bf Q} \cdot$) in Eq.~\ref{eq1}
with (${\bf Q} \times$). The transverse current spectra are also reported in
Fig.~2, where we observe the existence of a low energy excitation
($E \approx 4 $ meV at $Q=1.47$ nm$^{-1}$)
which disperses with a sound velocity of $\approx 4300 $ m/s
up to $\approx 15$ mev at $Q$=5 nm$^{-1}$.
The transverse modes become almost non-dispersing at $E \approx 20$ meV,
an energy that is definitely higher than that characteristic of the boson peak. 
Also the
propagating transverse dynamics may therefore contributes to the boson peak.
It is worth to note that, in the $Q$ range where the longitudinal and
transverse modes show a propagating character, the spectra $C_T(Q,E)$
and $C_L(Q,E)$ show a marked difference, again indicating the non-localized
nature of the vibrational modes. Indeed, for localized modes one should
not expect any difference between "longitudinal" and "transverse" dynamics,
being ${\bf Q}$ meaningless.

In conclusion, the numerical study of harmonic model of v-SiO$_2$
indicates that:
{\it i)}  The experimentally observed width of the excitations in the
$Q=1.4 \div 3.5$ nm$^{-1}$ range is due to structural effects, i.e. to
the fact that in this range $Q$ is no longer a good quantum number
in the topologically disordered silica.
Anharmonicity or other dynamical effects (interaction with two level systems,
soft potential modes, etc.) can only play a minor role.
{\it ii)} Both longitudinal and transverse modes are propagating in the
energy range covered by the boson peak:
their dispersion relations saturate at much higher energy.
{\it iii)} The explanation of the numerical data with a cross-over
model gives a $Q$ dependent cross-over energy which is 
inconsistent with the model itself.
{\it iv)} The transverse
and longitudinal dynamics keep their differences up to at least $Q=$ 5
nm$^{-1}$, which indicates the non-localized nature of the vibrational
states up to this $Q$-value.

We greatefully akwnowledge F. Sette for many discussions and suggestions,
P. Benassi and V. Mazzacurati for useful discussions, and
U. Buchenau and S.R.~Elliot for the preprints of ref.~\cite{BUCH3} and
\cite{Elliot2} respectively.

{\footnotesize{
\begin{center}
{\bf FIGURE CAPTIONS}
\end{center}

\begin{description}

\item  {
Fig.1 - $Q$ dependence of the inelastic neutron scattering
from vitreous silica at $1.3$ THz. 
The circle represent the data of ref.~\cite{BUCH3} at $T=1104$K, 
the full line the results of the present NMA analysis
multiplied by an arbitrary factor.
}

\item  {
Fig.2 - X-ray longitudinal (line plus open circles) 
and transverse (line plus solid circles) current spectra
obtained by the NMA at different $Q$ values.  
Open and solid up-triangles refer to longitudinal and transverse spectra 
respectively, obtained by MD ($216$ SiO$_2$ units at $300$ K).
Full lines are the DHO fits of NMA longitudinal current spectra.
}

\item  {
Fig.3 - a) Excitations energy $\Omega(Q)$ from the DHO model
for the IXS data at $T=1050$K (open circles) \cite{grenoble}
and for the present NMA on the $0$K configuration.
b) As in (a) but for the full width at half maximum
of the excitations $\Gamma(Q)$.
}

\item  {
Fig.4 - $Q$-dependence of the cross
over frequency $\omega_{IR}$ resulting from the fits of the
localised modes model of ref.~\cite{courtens} to the 
longitudinal current spectra, calculated from NMA. 
The inset shows an example of the fit quality.
}

\end{description}
}}
\end{document}